\documentclass[english]{cccconf}
\usepackage[comma,numbers,square,sort&compress]{natbib}
\usepackage{epstopdf}
\usepackage[T1]{fontenc}
\usepackage{amsfonts}
\usepackage{amsmath}
\usepackage{amssymb}
\usepackage{bm}
\usepackage{setspace}
\usepackage{algorithm}
\usepackage{algpseudocode}
\usepackage{listings}
\usepackage{graphicx}
\usepackage[caption=false,font=footnotesize]{subfig}
\usepackage{verbatim}
\usepackage{makecell}
\usepackage{booktabs}
\usepackage{ntheorem}
\usepackage{setspace}
\newtheorem{defn}{Definition}

\newtheorem{thm}{Theorem}

\newtheorem{assum}{Assumption}

\begin{document}

\title{Model-free Nearly Optimal Control of Constrained-Input Nonlinear Systems Based on Synchronous Reinforcement Learning}

\author{Han Zhao\aref{bupt},
        Lei Guo\aref{bupt}}



\affiliation[bupt]{School of Artificial Intelligence,
        Beijing University of Posts and Telecommunications, Beijing 100876, P.~R.~China
        \email{han{\_}zhao@bupt.edu.cn}}

\maketitle

\begin{abstract}
  In this paper a novel model-free algorithm is proposed. This algorithm can learn 
  the nearly optimal control law of constrained-input systems from online data without 
  requiring any \emph{a priori} knowledge of system dynamics. Based on the concept of 
  generalized policy iteration method, there are two neural networks (NNs), namely actor and 
  critic NN to approximate the optimal value function and optimal policy. The stability 
  of closed-loop systems and the convergence of weights are also guaranteed by Lyapunov 
  analysis.
\end{abstract}


\footnotetext{This work is supported by National Natural Science
Foundation (NNSF) of China under Grant 61105103.}

\section{Introduction}

The control method of nonlinear constraint systems has been widely researched 
and is a valuable practical problem. 
In the framework of optimal control, an optimal policy that minimize a user-defined performance index and the 
corresponding value function can be obtained by solving Hamilton-Jacobi-Bellman 
(HJB) equation, which is a partial differential equation in continuous-time nonlinear systems 
and the analytical solution is usually hard to find. Therefore, function 
approximation using neural networks (NNs) is a useful tool to solve HJB 
equation. These algorithms belong to the family of adaptive dynamic 
programming (ADP) or reinforcement learning (RL) methods.

ADP/RL methods have solved several discrete-time optimal control problems 
such as nonlinear zero-sum games \cite{Wei2018} and multi-agent systems \cite{Peng2019}, \cite{Peng2021}.
A considerable number of literatures have also focused on solving the 
continuous-time constrained-input optimal control problem via ADP/RL. 
In \cite{AbuKhalaf2005}, an off-line policy iteration (PI) algorithm that 
can iteratively solve HJB equation is proposed. The full knowledge of system 
dynamics and an initial stabilized controller are needed in this algorithm. 
In these years, many works are intended to relax the restriction of the 
\emph{a priori} knowledge of system dynamics. The concept of integral 
reinforcement learning (IRL, \cite{Vrabie2009}) and generalized PI 
\cite{Sutton1998} is brought in \cite{Modares2014a}, \cite{Modares2014b} to 
estimate the optimal value function in constrained-input problems. 
The initial stabilized controller is not needed in these methods, but the 
input gain matrix of the system is still used in policy improvement step.

Another way to solve this problem is to use an identifier NN. Xue \emph{et al.} 
\cite{Xue2020} proposed an algorithm based on actor-critic-identifier 
structure to estimate the system dynamics and solve HJB equation online. 
Comparing with other model-free methods, an extra identifier might bring 
additional burden on real-time computing.

In terms of IRL, Vamvoudakis \cite{Vamvoudakis2017} proposed a 
completely model-free Q-learning algorithm to solve continuous-time optimal control problem. 
Q-function combines the value 
function and the Hamiltonian in Pontryagin's minimum principle and provides 
the information of policy improvement without requiring any \emph{a priori}. 
Similarly, the restriction of the system input dynamics can be relaxed by adding 
exploration signal \cite{Lee2015}. Our previous work \cite{Guo2021} 
also present a novel model-free algorithm that combines the exploration signal 
and synchronous reinforcement learning to solve the adaptive optimal control 
problem in continuous-time nonlinear systems.

In this paper we present an algorithm to solve the continuous-time 
constrained-input optimal control problem without knowing any information 
of system dynamics or an initial stabilized controller. The identifier NN in 
\cite{Xue2020} is also not needed in this method. The remainder of this paper 
is organized as follows: Section \ref{sec2} provides the mathematic 
formulation of the optimal control for continuous-time systems with input 
constraints. In section \ref{sec3} we present the weights tuning law of 
actor-critic NNs based on exploration-HJB and synchronous reinforcement 
learning. In the process of learning, policy evaluation and improvement 
can be done simultaneously and no knowledge of system dynamics is required. 
The convergence of weights and stability of the closed-loop system is 
proved through Lyapunov analysis. Finally, we design two simulations to 
verify the effectiveness of our method.

\section{Problem Formulation}
\label{sec2}

Consider a continuous-time input-affine system
\begin{equation}
  \label{system}
  \dot x(t)=f(x(t))+g(x(t))u(t)\ \ \ x(0)=\xi,
\end{equation}
where $x \in \mathbb R^n$ and $u \in \mathbb R^m$ is the measurable 
state and input vector respectively. $\xi$ denotes the initial state of 
the system. $f(x) \in \mathbb R^n$ is called as the system drift dynamics and 
$g(x) \in \mathbb R^{n \times m}$ is the system input dynamics. The system 
dynamics $f(x)+g(x)u$ is assumed to be Lipschitz in a compact set $\Omega$ and 
satisfies $f(0)=0$.

The objective of optimal control is to design a controller that minimize a 
user-defined performance index. The performance index in this paper is defined 
as
\begin{equation}
  \label{index}
  J(x,u)=\int_0^\infty \left(Q(x(\tau))+U(u(\tau))\right){\rm d}\tau,
\end{equation}
both $Q(x)$ and $U(u)$ are positive definite functions.

In this paper, the constrained-input control problem is considered. The input 
vector needs to satisfy the constraint $\vert u_i \vert \le \lambda(i=1,...,m).$ 
To deal with this constraint, a non-quadratic integral cost function of input 
is usually employed \cite{Lyshevski1998}:

\begin{equation}
  \label{U}
  U(u)=2\int_0^u \lambda \bm{\vartheta}^{-1} (\frac{s}{\lambda})R {\rm d} \textit{s},
\end{equation}
where $\bm{\vartheta}(s)=\left[\vartheta(s_1),...,\vartheta(s_m)\right]^\top$ 
and $\bm{\vartheta}^{-1}(s)=\left[\vartheta^{-1}(s_1),...,\vartheta^{-1}(s_m)\right]^\top$. 
$\vartheta$ is a bounded one-to-one monotonic odd function. The hyperbolic 
tangent function $\vartheta(\cdot)=\tanh(\cdot)$ is often used. $R$ is a positive 
definite symmetric matrix. For the convenience of computing and analysis, 
$R$ is assumed to be diagonal, i.e. $R={\rm diag}(r_1,...,r_m).$

The goal is to find the optimal control law $u^*$ that can stabilize system 
\eqref{system} and minimize performance index \eqref{index}. The following value 
function is used:
\begin{equation}
  \label{VF}
  V^{\mu}=J(x,u)\vert_{u=\mu(x)},V^{\mu}(0)=0,
\end{equation}
where $\mu(x)$ is a feedback policy and $\mu(0)=0.$ The value function 
is well-defined only if the policy is admissible \cite{AbuKhalaf2005}.

\begin{defn}
  \label{admissiblepolicy}
  A constrained policy is admissible in $\Omega$, which is denoted as 
  $\mu \in \mathcal A(\Omega)$, if:
  1) The policy stabilize system \eqref{system} in $\Omega$;
  2) For any state $x \in \Omega, V^\mu(x)$ is finite;
  3) For any state $x \in \Omega, \vert \mu_i(x) \vert \le \lambda(i=1,...,m)$.
  $\Omega$ is called as the admissible domain and $\mathcal A(\Omega)$ is the 
  set of admissible policies.
\end{defn}

Assume that the set of admissible policy $\mathcal A(\Omega)$ of system 
\eqref{system} is not empty and $V^\mu \in \mathcal C^1(\Omega)$. It is 
obvious that there exists an optimal policy $\mu^*(x) \in \mathcal A(\Omega)$ 
so that

\begin{equation*}
  V^{\mu^*}(\xi)=\min_{\mu \in \mathcal A(\Omega)}\int_0^\infty \left(Q(x(\tau)+U(u(\tau)))\right){\rm d}\tau, \forall \xi \in \Omega.
\end{equation*}
The optimal value function satisfies $V^*(x)=V^{\mu^*}(x)$. In the remainder 
of the paper both of these two symbols may be used to describe the same 
variable.

According to the definition of the value function \eqref{VF}, the infinitesimal 
version of \eqref{index} is obtained as
\begin{equation}
  \label{Lyapunovfunction}
  0 = Q(x) + U(\mu(x)) + (\nabla V^\mu(x))^\top(f(x)+g(x)\mu(x)),
\end{equation}
where $\nabla V(x)=\partial V(x)/\partial x \in \mathbb R^n$. Define the 
Hamiltonian as
\begin{equation}
  \label{Hamiltonian}
  \mathcal H(x,u,\nabla V^\mu) = Q(x) + U(u) + (\nabla V^\mu)^\top(f(x)+g(x)u).
\end{equation}

For optimal value function $V^*$ and optimal control law $u^*$, the 
following Hamilton-Jacobi-Bellman (HJB) equation is satisfied:
\begin{equation}
  \label{HJB}
  \begin{aligned}
    &0 = Q(x) + U(u^*) + \nabla V^{*\top}(f(x)+g(x)u^*),\\
    &V^*(0) = 0.
  \end{aligned}
\end{equation}
The optimal policy can be derived by minimizing the Hamiltonian, for 
input-affine system \eqref{system} and the performance index \eqref{index} 
and \eqref{U}, the optimal policy can be directly denoted as
\begin{equation}
  \label{optimalpolicy}
  u^*(x)=\lambda {\bm \vartheta}\left(-\frac{1}{2\lambda}R^{-1}g^\top(x)\nabla V^*(x)\right).
\end{equation}
Let
\begin{equation}
  \label{optimalv}
  v^*(x)= -\frac{1}{2}R^{-1}g^\top(x)\nabla V^*(x),
\end{equation}
the optimal policy can be obtained as
\begin{equation}
  \label{optimalu}
  u^*(x)= \lambda {\bm \vartheta}\left(\frac{v^*(x)}{\lambda}\right).
\end{equation}

It is hard to solve the optimal value function and policy in \eqref{HJB} 
directly and the \emph{a priori} knowledge of system dynamics is required. 
Therefore, the online data-driven algorithm is a useful tool to relax the 
restriction.

\section{Online Synchronous Reinforcement Learning Algorithm to Solve HJB Equation}
\label{sec3}
In this section a data-driven method that learning the optimal policy is 
introduced. The framework of integral reinforcement learning (IRL) is proposed 
in \cite{Vrabie2009}. A completely model-free algorithm based on IRL is 
developed by Lee \emph{et al.} \cite{Lee2015} to solve the optimal control 
problem of input-affine systems. However, the initial stabilized controller is 
still required to start the iteration.

In our previous work \cite{Guo2021}, an algorithm called synchronous integral 
Q-learning is proposed, which can continuously update the weights of both two 
NNs to estimate the optimal value function and policy. In this paper we extend 
the algorithm to solve the constrained-input problem. By adding a bounded 
piece-wise continuous non-zero exploration signal $e$ into input, the system 
\eqref{system} is derived as 
\begin{equation}
  \label{expsystem}
  \dot x = f(x)+g(x)(u+e).
\end{equation}
Substituting \eqref{expsystem} into HJB equation \eqref{HJB}. For any time $t>T$ 
and time interval $T>0$, integrating \eqref{HJB} from $[t-T,t]$ along the 
trajectory of system \eqref{expsystem}, the following integral temporal 
difference equation is obtained as
\begin{equation}
  \label{ITD}
  \begin{aligned}
    V^*(x(t)) - V^*(x(t-T)) &- \int_{t-T}^t \nabla V^{*\top}(x)g(x)e(\tau) {\rm d}\tau\\
    &= -\int_{t-T}^t \left(Q(x)+U(u^*)\right){\rm d}\tau.
  \end{aligned}
\end{equation}
The following HJB equation is derived by substituting 
\eqref{optimalv}:
\begin{equation}
  \label{EHJB}
  \begin{aligned}
    V^*(x(t)) - V^*(x(t-T)) &+ 2\int_{t-T}^t v^{*\top}R e(\tau) {\rm d}\tau\\
    &= -\int_{t-T}^t \left(Q(x)+U(u^*)\right){\rm d}\tau.
  \end{aligned}
\end{equation}

Both $V^*(x)$ and $v^*(x)$ can be solved in integral exploration-HJB equation 
\eqref{EHJB} and the optimal policy is computed by \eqref{optimalu}. Note that 
the knowledge of system dynamics is not needed in these two equations.

Since the analytical solution of $V^*(x)$ and $v^*(x)$ cannot be determined. 
We can approximate them by choosing a proper structure of neural networks (NNs). 
Assume $V^*(x)$ is a smooth function. Based on the assumption of system 
dynamics, $v^*(x)$ is also smooth. Therefore, there exists two single-layer 
NNs, namely actor and critic NN, so that:

1) $V^*(x)$ and its gradient can be universal approximated:
\begin{equation}
  \label{approximateV}
  V^*(x)=w_c^\top \phi_c(x) + \varepsilon_c(x),
\end{equation}
\begin{equation}
  \label{approximateVgrad}
  \nabla V^*(x)= \nabla \phi_c^\top(x) w_c + \nabla \varepsilon_c(x),
\end{equation}

2) $v^*(x)$ can be universal approximated:
\begin{equation}
  \label{approximatev}
  \begin{aligned}
    v^*(x)&=-\frac{1}{2}R^{-1}g^\top(x) (\nabla \phi_c^\top(x) w_c + \nabla \varepsilon_c(x))\\
    &=w_a^\top \phi_a(x) + \varepsilon_a(x),
  \end{aligned}
\end{equation}
where $\phi_c(x):\mathbb R^n \to \mathbb R^{N_c},w_c,\varepsilon_c$ is the 
activation function, weights and the reconstruction error of the critic NN, 
respectively. $N_c$ is the number of neurons in the hidden layer of the critic 
NN. Similarly, in actor NN, we denote 
$\phi_a(x):\mathbb R^n \to \mathbb R^{N_a},w_a,\varepsilon_a$ is the 
activation function, weights and the reconstruction error, respectively. 
$N_a$ is the number of neurons in the hidden layer of the actor NN. 
$\varepsilon_c(x)$ and $\varepsilon_a(x)$ are bounded on a compact set $\Omega$. 
According to Weierstrass approximation theorem, when $N_c,N_a \to \infty$, we 
have $\varepsilon_c,\nabla \varepsilon_c, \varepsilon_a \to 0.$

Define the Bellman error $\varepsilon_B$ based on \eqref{EHJB} as
\begin{equation}
  \label{Bellmanerror}
  \begin{aligned}
    w_c^\top \Delta \phi_c &+ 2\int_{t-T}^t \phi_a^\top(x(\tau))w_aRe(\tau){\rm d}\tau\\
    &+\int_{t-T}^t \left(Q(x(\tau))+U(u^*(x(\tau))\right){\rm d}\tau = \varepsilon_B,
  \end{aligned}
\end{equation}
where $\Delta \phi_c = \phi_c(x(t))-\phi_c(x(t-T))$. The integral 
reinforcement is defined as
\begin{equation*}
  \rho(x,u)=\int_{t-T}^t \left(Q(x(\tau))+U(u(x(\tau))\right){\rm d}\tau.
\end{equation*}
Using the characteristic of Kronecker product, the following equation 
can be derived from \eqref{Bellmanerror}:
\begin{equation}
  \varepsilon_B - \rho(x,u^*) = W^\top \delta,
\end{equation}
where $W = [w_c^\top,{\rm col}\{w_a\}^\top]^\top$. ${\rm col}\{w_a\}$ is the reshaped column vector of $w_a$, and
\begin{equation*}
  \delta = [\Delta \phi_c^\top,(\int_{t-T}^t 2\phi_a \otimes (Re(\tau)){\rm d}\tau)^\top]^\top
\end{equation*}

Before introducing the weights tuning law, it is necessary to make some 
assumptions of NNs.

\begin{assum}
  \label{assum1}
  The system dynamics and NNs satisfy:

  a. $f(x)$ is Lipschitz and $g(x)$ is bounded.

  $\Arrowvert f(x) \Arrowvert \le b_f\Arrowvert x \Arrowvert,\Arrowvert g(x) \Arrowvert \le b_g.$
  
  b. The reconstruction errors and the gradient of critic NN's error is bounded.
  
  $\Arrowvert \varepsilon_c \Arrowvert \le b_{\varepsilon c},\Arrowvert \varepsilon_a \Arrowvert \le b_{\varepsilon a},\Arrowvert \nabla \varepsilon_c \Arrowvert \le b_{\varepsilon cx}.$
  
  c. The activation functions and the gradient of critic NN's activation function is bounded.
  
  $\Arrowvert \phi_c(x) \Arrowvert \le b_{\phi c},\Arrowvert \phi_a(x) \Arrowvert \le b_{\phi a},\Arrowvert \nabla \phi_c(x) \Arrowvert \le b_{\phi cx}.$

  d. The optimal weights of NNs $\Arrowvert w_c \Arrowvert,\Arrowvert w_a \Arrowvert$ 
  are bounded, and the amplitude of exploration signal is bounded by $b_e$.
\end{assum}

These assumptions are trivial and can be easily satisfied except the assumption 
of $g(x).$ With the assumption that system dynamics is Lipschitz, Bellman error 
$\varepsilon_B$ is bounded on a compact set. When $N_c,N_a \to \infty$, 
$\varepsilon_B \to 0$ universally \cite{AbuKhalaf2005}.

In this paper we use $\vartheta(\cdot)=\tanh(\cdot)$. The non-quadratic cost 
$U(u)$ is derived as
\begin{equation}
  \label{nointU}
  \begin{aligned}
    U(u)&=2\int_0^u \lambda \tanh^{-1}(\frac{s}{\lambda})R{\rm d}s\\
    &=2\lambda u^\top R \tanh^{-1}(\frac{u}{\lambda}) + \lambda^2 \overline R \ln \left(\bm 1_m - (\frac{u}{\lambda})^2\right),
  \end{aligned}
\end{equation}
where $\overline R = [r_1,...,r_m], \bm 1_m$ is the $m$-dimensional vector with 
all elements are one. According to \eqref{Bellmanerror} and \eqref{nointU}, the 
approximate exploration-HJB equation is obtained as
\begin{equation}
  \label{approximateEHJB}
  \begin{aligned}
    &\int_{t-T}^t \left( -Q - 2\lambda \tanh^\top (\frac{w_a^\top \phi_a}{\lambda})Rw_a^\top \phi_a \right.\\
    &\left. -\lambda^2 \overline R \ln\left(\bm 1_m - \tanh^2 (\frac{w_a^\top \phi_a}{\lambda}) \right) + \varepsilon_{HJB} \right) {\rm d}\tau = W^\top \delta,
  \end{aligned}
\end{equation}
in which $\varepsilon_{HJB}(x)$ is the error raised from the reconstruction 
errors of two NNs. Since the ideal weights $w_c$ and $w_a$ are unknown, they 
are approximated in real time as
\begin{equation}
  \label{estimateV}
  \hat V(x)=\hat w_c^\top \phi_c(x),
\end{equation}
\begin{equation}
  \label{estimatev}
  \hat v_1(x)=\hat w_{a1}^\top \phi_a(x).
\end{equation}
The optimal policy is obtained as
\begin{equation*}
  u(x)= \lambda \tanh \left(\frac{w_a^\top \phi_a(x) + \varepsilon_a(x)}{\lambda}\right).
\end{equation*}
We use an actor NN to approximate the optimal policy as
\begin{equation}
  \label{estimateu}
  \hat u(x)=\lambda \tanh \left(\frac{\hat v_2(x)}{\lambda}\right),
\end{equation}
where $\hat v_2(x) = \hat w_{a2}^\top \phi_a(x)$. Note that $\hat v_1$ and 
$\hat v_2$ have the same approximate structure. However, the estimated weights 
in \eqref{estimatev} will not be directly implemented on the controller for 
Lyapunov stability. Define $\hat W = [\hat w_c^\top, {\rm col}\{\hat w_{a1}\}]^\top$, 
the approximate Bellman error of the critic NN is obtained as
\begin{equation}
  \label{appBellmanerror}
  E = \hat W^\top \delta + \rho(x,\hat u).
\end{equation}

In order to minimize the error, we define the objective function of critic NN as $K = \frac{1}{2}E^\top E.$ 
The modified gradient-descent law is obtained as
\begin{equation}
  \label{critictuninglaw}
  \dot {\hat W} = -\alpha_1 \frac{\delta}{1+\delta^\top \delta}E,
\end{equation}
where $\alpha_1$ is the learning rate that determines the speed of convergence. In 
order to guarantee the stability of the closed-loop system and the convergence. 
Define the approximate error of the actor NN as
\begin{equation}
  \label{appactorerror}
  E_u = \lambda R \left(\tanh(\frac{\hat w_{a2}^\top \phi_a}{\lambda})-\tanh(\frac{\hat w_{a1}^\top \phi_a}{\lambda})\right)
\end{equation}
The gradient-descent tuning law of $\hat w_{a2}$ is set as
\begin{equation}
  \label{actortuninglaw}
  \begin{aligned}
    {\rm col}\{\dot {\hat w}_{a2}\} = &-\alpha_2 \bigg(Y {\rm col}\{\hat w_{a2}\}\\
    &+ E_u \otimes \phi_a - \Bigl(E_u \tanh^2(\frac{\hat w_{a2}^\top \phi_a}{\lambda})\Bigr) \otimes \phi_a \bigg),
  \end{aligned}
\end{equation}
where $Y$ is a designed parameter to guarantee stability. Define 
$m_s=1+\delta^\top \delta$ and $\overline \delta = \delta/m_s$, the following 
theorem is derived.

\begin{thm}
  \label{thm1}
  If all terms in Assumption \ref{assum1} is satisfied and signal
  $\overline \delta(t)$ is persistently excited \cite{Ioannou2006}, i.e. 
  there exists two constant $\beta_1<\beta_2$ such that
  $\beta_1 I \le \int_{t-T}^t \overline \delta^\top \overline \delta {\rm d}\tau \le \beta_2 I, \forall t>T.$
  There exists a positive integer $N_0$ and a sufficiently small reinforcement 
  interval $T>0$ such that, when the number of neurons $N_c,N_a > N_0$, the 
  states in closed-loop system, the estimated error of $\hat V(x),\hat v_1(x)$ 
  and $\hat u(x)$ are uniformly ultimately bounded.
\end{thm}

\section{Proof of Theorem \ref{thm1}}
  Define the error of weights as $\tilde w_c = w_c - \hat w_c, \tilde w_{a1} = w_a - \hat w_{a1}$ 
  and $\tilde w_{a2} = w_a - \hat w_{a2}$. Consider the following Lyapunov 
  candidate
  \begin{equation}
    \label{Lyapunov}
    L = V^*(x(t)) + \frac{1}{2}\tilde W^\top(t) \alpha_1^{-1} \tilde W(t) + \frac{1}{2}\tilde w_{a2}^\top(t) \alpha_2^{-1} \tilde w_{a2}(t).
  \end{equation}
  The derivative of \eqref{Lyapunov} to time is
  \begin{equation}
    \label{dLyapunov}
    \dot L = \dot V^*(x(t)) + \tilde W^\top(t) \alpha_1^{-1} \dot {\tilde W}(t) + \tilde w_{a2}^\top(t) \alpha_2^{-1} \dot {\tilde w}_{a2}(t).
  \end{equation}

  The first term in \eqref{dLyapunov} is obtained as
  \begin{equation}
    \label{dotV1}
    \begin{aligned}
      \dot V^* =\  &w_c^\top \nabla \phi_c (f + g \lambda \tanh(\frac{\hat w_{a2}^\top \phi_a}{\lambda}) + ge)\\
      &+\nabla \varepsilon_c^\top (f + g \lambda \tanh(\frac{\hat w_{a2}^\top \phi_a}{\lambda}) + ge)
    \end{aligned}
  \end{equation}
  From \eqref{approximatev}, we have
  \begin{equation}
    \label{32}
    \begin{aligned}
      &w_c^\top \nabla \phi_c g \lambda \tanh(\frac{\hat w_{a2}^\top \phi_a}{\lambda}) = -2\phi_a^\top w_a R \lambda \tanh(\frac{\hat w_{a2}^\top \phi_a}{\lambda})\\
      &-2\varepsilon_a^\top R \lambda \tanh(\frac{\hat w_{a2}^\top \phi_a}{\lambda} - \nabla \varepsilon_c^\top g \lambda \tanh(\frac{\hat w_{a2}^\top \phi_a}{\lambda}).
    \end{aligned}
  \end{equation}
  
  Using $w_a = \hat w_{a2} + \tilde w_{a2}$ and the fact that 
  $x^\top \tanh(x) \ge 0, \forall x$ for the first term on the right side of 
  \eqref{32}, it can be obtained as
  \begin{equation}
    -2\phi_a^\top w_a R \lambda \tanh(\frac{\hat w_{a2}^\top \phi_a}{\lambda}) \le -2\phi_a^\top \tilde w_{a2}^\top R \lambda \tanh(\frac{\hat w_{a2} \phi_a}{\lambda}).
  \end{equation}
  Therefore, \eqref{dotV1} turns into
  \begin{equation}
    \label{dotV2}
    \begin{aligned}
      &\dot V^* \le w_c^\top \nabla \phi_c (f + g\lambda \tanh(\frac{\hat w_{a2}^\top \phi_a}{\lambda})+ge) + \varepsilon_1\\
      &-2\phi_a^\top w_a R \lambda \tanh(\frac{\hat w_{a2}^\top \phi_a}{\lambda}) - w_c^\top \nabla \phi_c g\lambda \tanh(\frac{\hat w_{a2}^\top \phi_a}{\lambda}),\\
    \end{aligned}
  \end{equation}
  where $\varepsilon_1 = \nabla \varepsilon_c^\top (f+ge) - 2\varepsilon_a^\top R \lambda \tanh(\hat w_{a2}^\top \phi_a / \lambda)$. 
  According to assumption \ref{assum1}, it satisfies
  \begin{equation}
    \label{eps1}
    \varepsilon_1 \le b_{\varepsilon cx}b_f \Arrowvert x \Arrowvert + b_{\varepsilon cx} b_g b_e + 2\lambda b_{\varepsilon a}\sigma_{\min}(R),
  \end{equation}
  where $\sigma_{\min}(R)$ denotes the minimum singular value of matrix $R$. 
  Substituting the derivative of the approximate HJB equation \eqref{approximateEHJB} 
  into \eqref{dotV2}, we have
  \begin{equation}
    \label{dotV3}
    \begin{aligned}
      &\dot V^* \le -Q -U\left( \tanh(\frac{w_a^\top \phi_a}{\lambda}) \right) + \varepsilon_{HJB} + \varepsilon_1\\
      &-2\phi_a^\top w_a R \lambda \tanh(\frac{\hat w_{a2}^\top \phi_a}{\lambda}) - w_c^\top \nabla \phi_c g\lambda \tanh(\frac{\hat w_{a2}^\top \phi_a}{\lambda}).\\
    \end{aligned}
  \end{equation}
  The last term in \eqref{dotV3} satisfies
  \begin{equation}
    -w_c^\top \nabla \phi_c g \lambda \tanh(\frac{w_a^\top \phi_a}{\lambda}) \le \lambda b_g b_{\phi cx} \Arrowvert w_c \Arrowvert.
  \end{equation}

  Since $Q$ and $U$ are positive definite, there exists a $q>0$ such that 
  $x^\top q x < Q+U$. Therefore, \eqref{dotV3} can be derived as
  \begin{equation}
    \label{dotV4}
    \dot V^* \le -x^\top q x + k_1 \Arrowvert x \Arrowvert + k_2 - 2\phi_a^\top \tilde w_{a2} R \lambda \tanh(\frac{\hat w_{a2}^\top \phi_a}{\lambda}),
  \end{equation}
  where
  \begin{equation*}
    \begin{aligned}
      &k_1 = b_{\varepsilon cx} b_f,\\
      &k_2 = b_{\varepsilon cx} b_g b_e + 2\lambda b_{\varepsilon a} \sigma_{\min}(R) + \lambda b_g b_{\phi cx}\Arrowvert w_c \Arrowvert + \varepsilon_h,
    \end{aligned}
  \end{equation*}
  $\varepsilon_h$ is the upper bound of $\varepsilon_{HJB}$.

  Now analyzing the second term in \eqref{dLyapunov}. According to the tuning law 
  \eqref{critictuninglaw}, we have the following error dynamics
  \begin{equation}
    \label{errordynamics}
    \begin{aligned}
    &\dot {\tilde W}(t) = \alpha_1 (\overline \delta / m_s) E(t)\\
    &y(t) = \overline \delta^\top \tilde W(t)
    \end{aligned}
  \end{equation}
  Substituting HJB equation \eqref{approximateEHJB} into \eqref{errordynamics}, 
 one has
  \begin{equation}
    \label{E}
    \begin{aligned}
      E(t) &= \hat W^\top(t) \delta(t) + \int_{t-T}^t (Q+ \hat U){\rm d}\tau\\
      &=\int_{t-T}^t \Big(Q+U(\hat u)-Q-U(u^*)+ \varepsilon_{HJB}\\
      &+\hat w_c^\top (t) \nabla \phi_c (f + g\lambda \tanh(\frac{\hat w_{a2}^\top \phi_a}{\lambda})+ge)\\
      &-w_c^\top \nabla \phi_c (f + g\lambda \tanh(\frac{w_a^\top \phi_a}{\lambda}) +ge)\Big){\rm d}\tau\\
      &- 2{\rm col}\{\tilde w_{a1}\}^\top \int_{t-T}^t \phi_a \otimes Re {\rm d}\tau,
    \end{aligned}
  \end{equation}
  According to the definition of $U$, we have
  \begin{equation}
    \label{tildeU}
    \begin{aligned}
      &U(\hat u)-U(u^*) =\\
      &2\lambda \phi_a^\top \hat w_{a2} R \tanh(\frac{\hat w_{a2}^\top \phi_a}{\lambda}) -2\lambda \phi_a^\top w_a R \tanh(\frac{w_a^\top \phi_a}{\lambda})\\
      &+ \lambda^2 \overline R \left(\ln \left( \bm 1_m - \tanh^2 (\frac{\hat w_{a2}^\top \phi_a}{\lambda}) \right)\right)\\
      &- \lambda^2 \overline R \left(\ln \left( \bm 1_m - \tanh^2 (\frac{\hat w_a^\top \phi_a}{\lambda}) \right)\right)
    \end{aligned}
  \end{equation}
  According to \cite{Modares2014a}, the third term in \eqref{tildeU} 
  can be written as
  \begin{equation}
    \label{ln1}
    \begin{aligned}
    &\ln \left(\bm 1_m - \tanh^2 (\frac{\hat w_{a2}^\top \phi_a}{\lambda})\right)\\
    &= \ln 4 -2\frac{\hat w_{a2}^\top \phi_a}{\lambda}{\rm sgn}(\frac{\hat w_{a2}^\top \phi_a}{\lambda}) + \varepsilon_{\hat D},
    \end{aligned}
  \end{equation}
  where $\Arrowvert \varepsilon_{\hat D} \Arrowvert \le \ln 4$. Similarly, we have
  \begin{equation}
    \label{ln2}
    \begin{aligned}
    &\ln \left(\bm 1_m - \tanh^2 (\frac{w_a^\top \phi_a}{\lambda})\right)\\
    &= \ln 4 -2\frac{w_a^\top \phi_a}{\lambda}{\rm sgn}(\frac{w_a^\top \phi_a}{\lambda}) + \varepsilon_D,
    \end{aligned}
  \end{equation}
  where $\Arrowvert \varepsilon_D \Arrowvert \le \ln 4$. The following inequality 
  is derived from \eqref{tildeU}:
  \begin{equation}
    \label{tildeU2}
    \begin{aligned}
      &U(\hat u)-U(u^*) = \lambda^2 \overline R (\varepsilon_{\hat D} - \varepsilon_D)\\
      &2\lambda \phi_a^\top \hat w_{a2} R \tanh(\frac{\hat w_{a2}^\top \phi_a}{\lambda}) -2\lambda \phi_a^\top w_a R \tanh(\frac{w_a^\top \phi_a}{\lambda})\\
      &-2\lambda^2 \overline R \hat w_{a2}^\top \phi_a \ {\rm sgn}(\frac{\hat w_{a2}^\top \phi_a}{\lambda}) +2\lambda^2 \overline R w_a^\top \phi_a\ {\rm sgn}(\frac{w_a^\top \phi_a}{\lambda}).\\
    \end{aligned}
  \end{equation}

  In \cite{Modares2014a}, a continuous approximation of $x\ {\rm sgn}(x)$ is provided:
  \begin{equation}
    \label{approxisgn}
    0 \le x\ {\rm sgn}(x) - x \tanh(\kappa x) \le \frac{3.59}{\kappa}.
  \end{equation}
  Combining \eqref{tildeU2} and \eqref{approxisgn}, the approximate HJB error is
  \begin{equation}
    \label{E2}
    \begin{aligned}
      E(t) &= \int_{t-T}^t \Big( 2 \lambda \phi_a^\top \hat w_{a2} R \tanh(\frac{\hat w_{a2}^\top \phi_a}{\lambda}) \\
      &- 2 \lambda \phi_a^\top w_a R \tanh(\frac{w_a^\top \phi_a}{\lambda}) + \varepsilon_\kappa + \varepsilon_{HJB}\\
      &+ 2 \lambda^2 \overline R w_a^\top \phi_a \left(\tanh(\kappa \frac{w_a^\top \phi_a}{\lambda})- \tanh(\kappa \frac{\hat w_{a2}^\top \phi_a}{\lambda})\right)\\
      &+ 2 \lambda^2 \overline R \tilde w_{a2}^\top \phi_a \tanh(\kappa \frac{\hat w_{a2}^\top \phi_a}{\lambda}) + \lambda^2 \overline R (\varepsilon_{\hat D} - \varepsilon_D)\\
      &+\hat w_c^\top(t) \nabla \phi_c (f + ge) + \hat w_c^\top(t) \nabla \phi_c g \lambda \tanh(\frac{\hat w_{a2}^\top \phi_a}{\lambda})\\
      &- w_c^\top \nabla \phi_c (f + ge) - w_c^\top \nabla \phi_c g \lambda \tanh(\frac{w_a^\top \phi_a}{\lambda}) \Big){\rm d}\tau\\
      &- 2 {\rm col}\{w_{a1}(t)\}^\top \int_{t-T}^t \phi_a \otimes Re {\rm d}\tau,
    \end{aligned}
  \end{equation}
  in which the approximate error satisfies $0 \le \varepsilon_\kappa \le 7.18 / \kappa$. 
  
  Substituting \eqref{32} into \eqref{E2}, we can get
  \begin{equation}
    \label{E3}
      E(t) = -\overline \delta ^\top \tilde W(t) + \int_{t-T}^t {\rm col}\{\tilde w_{a2}(\tau)\}^\top M {\rm d}\tau + N(t),
  \end{equation}
  where
  \begin{equation*}
    \begin{aligned}
      &M = 2 \lambda \phi_a \otimes R \left(\lambda \tanh(\kappa \frac{\hat w_{a2}^\top \phi_a}{\lambda}) - \tanh(\frac{\hat w_{a2}^\top \phi_a}{\lambda})\right),\\
      &N(t) = \int_{t-T}^t \Big( \varepsilon_{HJB} + \lambda^2 \overline R (\varepsilon_{\hat D}- \varepsilon_D) + \varepsilon_\kappa + \varepsilon_2\\
      &+2 \lambda^2 \overline R w_a^\top \phi_a \big(\tanh(\frac{\kappa w_a^\top \phi_a}{\lambda}) - \tanh(\kappa \frac{\hat w_{a2}^\top \phi_a}{\lambda})\big)\Big){\rm d} \tau,\\
    \end{aligned}
  \end{equation*}
  and
  \begin{equation*}
    \varepsilon_2 = (2\varepsilon_a^\top R + \nabla \varepsilon_c^\top g) \lambda \left(\tanh(\frac{w_a^\top \phi_a}{\lambda}) - \tanh(\frac{\hat w_{a2}^\top \phi_a}{\lambda})\right).
  \end{equation*}
  Substituting \eqref{E3} into \eqref{errordynamics}, the error dynamics can be rewritten as
  \begin{equation}
    \label{errordynamics2}
    \begin{aligned}
    \dot {\tilde W}(t) = \alpha_1 \Big(&-\overline \delta \overline \delta^\top \tilde W(t) \\
    &+ \frac{\overline \delta}{m_s} \int_{t-T}^t {\rm col}\{\tilde w_{a2}\}^\top M {\rm d}\tau + \frac{\overline \delta}{m_s}N(t) \Big).
    \end{aligned}
  \end{equation}
  The second term in \eqref{dLyapunov} is obtained as
  \begin{equation}
    \label{dotW}
    \begin{aligned}
    &\tilde W^\top(t) \alpha_1^{-1} \dot {\tilde W}(t) = -\tilde W^\top(t) \overline \delta \overline \delta^\top \tilde W(t) \\
    &+ \tilde W^\top(t) \frac{\overline \delta}{m_s} \int_{t-T}^t {\rm col}\{\tilde w_{a2}\}^\top M {\rm d}\tau + \tilde W^\top(t) \frac{\overline \delta}{m_s}N(t).
    \end{aligned}
  \end{equation}
  For a sufficiently small reinforcement interval, the integral term in 
  \eqref{dotW} can be approximated as
  \begin{equation}
    \label{appint}
    \int_{t-T}^t {\rm col}\{\tilde w_{a2}(\tau)\}^\top M {\rm d}\tau \approx T M^\top {\rm col}\{\tilde w_{a2}(t)\}.
  \end{equation}

  Substituting \eqref{appint} into \eqref{dotW} and applying Young's inequality, then \eqref{dotW} is derived as
  \begin{equation}
    \label{dotW3}
    \begin{aligned}
    \tilde W^\top(t) \alpha_1^{-1} \dot {\tilde W}(t) & \le -d \tilde W^\top(t) \overline \delta \overline \delta^\top \tilde W(t) + \tilde W^\top(t) \frac{\overline \delta}{m_s}N(t)\\
    &+\frac{\varepsilon}{2m_s^2} {\rm col}\{\tilde w_{a2}(t)\}^\top M M ^\top {\rm col}\{\tilde w_{a2}(t)\},
    \end{aligned}
  \end{equation}
  where $d = 1 - T^2/2\varepsilon$. Eq. \eqref{dLyapunov} can be obtained as
  \begin{equation}
    \label{dL2}
    \begin{aligned}
    \dot L \le & -x^\top q x + k_1 \Arrowvert x \Arrowvert + k_2\\
    &-d \tilde W^\top(t) \overline \delta \overline \delta^\top \tilde W(t) + \tilde W^\top(t) \frac{\overline \delta}{m_s}N(t)\\
    &+\frac{\varepsilon}{2m_s^2} {\rm col}\{\tilde w_{a2}\}^\top M M ^\top {\rm col}\{\tilde w_{a2}\}\\
    &- 2\phi_a^\top \tilde w_{a2} R \lambda \tanh(\frac{\hat w_{a2}^\top \phi_a}{\lambda})\\
    &+ {\rm col}\{\tilde w_{a2}\}^\top \alpha_2^{-1} {\rm col}\{\dot {\tilde w}_{a2}\}.
    \end{aligned}
  \end{equation}
  Substituting the tuning law of $\hat w_{a2}$ \eqref{actortuninglaw} and $\hat w_{a2} = w_a - \tilde w_{a2}$, 
  the last two term of \eqref{dL2} can be written as
  \begin{equation}
    \label{dwa2}
    \begin{aligned}
    &{\rm col}\{\tilde w_{a2}\}^\top \alpha_2^{-1} {\rm col}\{\dot {\tilde w}_{a2}\}- 2\phi_a^\top \tilde w_{a2} R \lambda \tanh(\frac{\hat w_{a2}^\top \phi_a}{\lambda})\\
    &=-{\rm col}\{\tilde w_{a2}\}^\top Y {\rm col}\{\tilde w_{a2}\} + {\rm col}\{\tilde w_{a2}\}^\top k_3,
    \end{aligned}
  \end{equation}
  where
  \begin{equation*}
    \begin{aligned}
    &k_3 = Y {\rm col}\{w_a\} - \Bigl(E_u \tanh^2(\frac{\hat w_{a2}^\top \phi_a}{\lambda})\Bigr) \otimes \phi_a \\
    &- \lambda R \tanh(\frac{\hat w_{a2}^\top \phi_a}{\lambda}) \otimes \phi_a -\lambda R \tanh(\frac{\hat w_{a1}^\top \phi_a}{\lambda}) \otimes \phi_a
    \end{aligned}
  \end{equation*}
  
  Eq. \eqref{dL2} becomes
  \begin{equation}
    \label{dL3}
    \begin{aligned}
    \dot L \le & -x^\top q x + k_1 \Arrowvert x \Arrowvert + k_2\\
    &-d \tilde W^\top(t) \overline \delta \overline \delta^\top \tilde W(t) + \tilde W^\top(t) \frac{\overline \delta}{m_s}N(t)\\
    &-{\rm col}\{\tilde w_{a2}\}^\top B {\rm col}\{\tilde w_{a2}\} + {\rm col}\{\tilde w_{a2}\}^\top k_3,
    \end{aligned}
  \end{equation}
  where $B = Y - \varepsilon M M^\top/ 2m_s^2$. Choose $T$ and $Y$ properly such 
  that both $d$ and $B$ are positive. Thus, $\dot L$ is negative if
  \begin{equation}
    \Arrowvert x \Arrowvert > \frac{k_1}{2\sigma_{\min}(q)}+\sqrt{\frac{k_1^2}{4\sigma_{\min}^2(q)}+\frac{k_2}{\sigma_{\min}(q)}},
  \end{equation}
  \begin{equation}
    \Arrowvert \overline \delta^\top \tilde W \Arrowvert > \frac{N}{d},
  \end{equation}
  \begin{equation}
    \Arrowvert \tilde w_{a2} \Arrowvert > \frac{k_3}{\sigma_{\min}(B)}.
  \end{equation}
  The inequalities above shows that the states of the closed-loop system, the 
  output of the error dynamics and the error $\tilde w_{a2}$ are UUB. Since the 
  signal $\overline \delta(t)$ is persistently excited. The weights of critic NN
  are also UUB.

\section{Simulation}
In this section we design two experiments to verify the effectiveness of our 
method. Since there are no analytic solutions to nonlinear optimal 
constrained-input control problem, we choose our first case as a linear system 
with a large enough upper bound that the input will not be near to the saturation. 
The optimal solution in this case should be same as the standard linear 
quadratic regulator (LQR) problem.

\subsection{Case 1: Linear System}
The system in the first case is chosen as
\begin{equation}
  \dot x = \left[\begin{matrix} 1 & 0 \\ 0 & -2\end{matrix}\right]x + \left[\begin{matrix} 2 \\ 1\end{matrix}\right]u.
\end{equation}
The cost function is defined as $Q=x_1^2+x_2^2$ and $R=1$. The input constraint 
is $\lambda=30.$ It becomes a LQR problem near the origin and the optimal value 
function is quadratic in the states and the optimal control law is linear. 
Therefore, we choose the activation functions as $\phi_c = [x_1^2,x_1x_2,x_2^2]^\top$ 
and $\phi_a = [x_1,x_2]^\top$. By solving algebraic Riccati equation, the optimal 
weights are obtained as $W^*=[0.8779,-0.1904,0.2492,-1.6601,-0.0577]^\top.$ As 
for the hyperparameters, we set reinforcement interval as $T=0.01\ \rm s$ and 
the learning rate of two NNs is set as $\alpha_1=10^3$ and $\alpha_2=20$, 
respectively. The exploration signal we add into the input is 
$e=\sum_{k=1}^{100}\sin (\omega_k t)$, where $\omega_k$ is uniformly sampled from 
$[-50,50]$. 

After $180\ \rm s$, the exploration signal is removed and the simulation 
end at $t_f = 200\ \rm s$. The weights converge to 
$\hat W(t_f) = [0.8777, -0.1876, 0.2461, -1.6618, -0.0589]^\top$ and 
$\hat w_{a2}(t_f) = [-1.6618, -0.0596]^\top$, which are close to the optimal 
weights.

\subsection{Case 2: Nonlinear System}
The second case is a nonlinear system \eqref{system} with 
$f(x) = [-x_1+x_2,-0.5(x_1+x_2)+0.5x_2(\cos(2x_1)+2)^2]^\top$ and 
$g(x) = [0,\cos(2x_1)+2]^\top$. In this case the upper bound of input is set as 
$\lambda=0.5$, we choose 
$\phi_c = [x_1^2,x_2^2,x_1x_2,x_1^4,x_2^4, x_1^3x_2,x_1^2x_2^2,x_1x_2^3]^\top$ 
and $\phi_a = [x_1,x_2,x_1^2,x_2^2,x_1x_2,x_1^3,x_2^3,x_1^2x_2,x_1x_2^2]^\top$. 
The learning rates are set as $\alpha_1 = 10^5, \alpha_2 = 10.$ Due to the 
input constraint, the exploration signal is chosen as
\begin{equation*}
  e(t) = \lambda \tanh (\frac{\frac{1}{30}\sum_{k=1}^{100}\sin(\omega_k t)+\hat w_{a2}^\top(t)\phi_a}{\lambda})-u(t).
\end{equation*}
After learning, the weights converge to
\begin{equation*}
  \begin{aligned}
    \hat w_c(t_f)=[0.0601,&1.0401,-0.1214,-0.0755,\\
    &0.1327,-0.0816,-0.0551,0.0048]^\top,
  \end{aligned}
\end{equation*}
\begin{equation*}
  \begin{aligned}
    \hat w_{a1}(t_f)=[0.&1876,-3.1325,-0.0278,-0.2763,-0.2390,\\
    &-2.0274,-0.0607,1.9738, 0.6572]^\top,
  \end{aligned}
\end{equation*}
\begin{equation*}
  \begin{aligned}
    \hat w_{a2}(t_f)=[0.&1826,-3.1356,-0.0164,-0.5104,-0.2490,\\
    &-2.0156,-0.0420,1.9659, 0.6565]^\top.
  \end{aligned}
\end{equation*}

\section{Conclusion}
This paper presents a novel adaptive optimal control method to solve 
constrained-input problem in a completely model-free way. By adding 
exploration signal, actor and critic NNs can simultaneously update 
their weights and no \emph{a priori} knowledge of system dynamics or 
an initial admissible policy is required during the learning phase. 
The efficacy of the proposed method is also shown in simulation results.

\bibliographystyle{ieeetr} 
\bibliography{sample_english}

\begin{thebibliography}{10}

\bibitem{Wei2018}
Q.~Wei, D.~Liu, Q.~Lin, and R.~Song, ``Adaptive dynamic programming for
  discrete-time zero-sum games,'' {\em IEEE Transactions on Neural Networks and
  Learning Systems}, vol.~29, no.~4, pp.~957--969, 2018.

\bibitem{Peng2019}
Z.~Peng, Y.~Zhao, J.~Hu, and B.~K. Ghosh, ``Data-driven optimal tracking
  control of discrete-time multi-agent systems with two-stage policy iteration
  algorithm,'' {\em Information Sciences}, vol.~481, pp.~189--202, 2019.

\bibitem{Peng2021}
Z.~Peng, Y.~Zhao, J.~Hu, R.~Luo, B.~K. Ghosh, and S.~K. Nguang,
  ``Input–output data-based output antisynchronization control of multiagent
  systems using reinforcement learning approach,'' {\em IEEE Transactions on
  Industrial Informatics}, vol.~17, no.~11, pp.~7359--7367, 2021.

\bibitem{AbuKhalaf2005}
M.~Abu-Khalaf and F.~Lewis, ``Nearly optimal control laws for nonlinear systems
  with saturating actuators using a neural network {HJB} approach,'' {\em
  Automatica}, vol.~41, pp.~779--791, 2005.

\bibitem{Vrabie2009}
D.~Vrabie, O.~Pastravanu, F.~Lewis, and M.~Abu-Khalaf, ``Adaptive optimal
  control for continuous-time linear systems based on policy iteration,'' {\em
  Automatica}, vol.~45, pp.~477--484, 2009.

\bibitem{Sutton1998}
R.~Sutton and A.~Barto, {\em Reinforcement learning: an introduction}.
\newblock Cambridge University Press, 1998.

\bibitem{Modares2014a}
H.~Modares, F.~Lewis, and M.-B. Naghibi-Sistani, ``Integral reinforcement
  learning and experience replay for adaptive optimal control of
  partially-unknown constrained-input continuous-time systems,'' {\em
  Automatica}, vol.~50, no.~1, pp.~193--202, 2014.

\bibitem{Modares2014b}
H.~Modares and F.~Lewis, ``Optimal tracking control of nonlinear
  partially-unknown constrained-input systems using integral reinforcement
  learning,'' {\em Automatica}, vol.~50, no.~7, pp.~1780--1792, 2014.

\bibitem{Xue2020}
S.~Xue, B.~Luo, D.~Liu, and Y.~Yang, ``Constrained event-triggered ${H}_\infty$
  control based on adaptive dynamic programming with concurrent learning,''
  {\em IEEE Transactions on Systems, Man, and Cybernetics: Systems}, pp.~1--13,
  2020.

\bibitem{Vamvoudakis2017}
K.~Vamvoudakis and F.~Lewis, ``Q-learning for continuous-time linear systems: a
  model-free infinite horizon optimal control approach,'' {\em Systems \&
  Control Letters}, vol.~100, pp.~14--20, 2017.

\bibitem{Lee2015}
J.~Lee, J.~Park, and Y.~Choi, ``Integral reinforcement learning for
  continuous-time input-affine nonlinear systems with simultaneous invariant
  explorations,'' {\em IEEE Transactions on Neural Networks and Learning
  Systems}, vol.~26, pp.~916--932, 2015.

\bibitem{Guo2021}
L.~Guo and H.~Zhao, ``Online adaptive optimal control algorithm based on
  synchronous integral reinforcement learning with explorations,'' {\em arXiv
  e-prints}, 2021.
\newblock arxiv.org/abs/2105.09006.

\bibitem{Lyshevski1998}
S.~Lyshevski, ``Optimal control of nonlinear continuous-time systems: design of
  bounded controllers via generalized nonquadratic functionals,'' in {\em
  Proceedings of the 1998 American Control Conference}, vol.~1, pp.~205--209,
  1998.

\bibitem{Ioannou2006}
P.~Ioannou and B.~Fidan, {\em Adaptive control tutorial}.
\newblock SIAM, 2006.

\end{thebibliography}

\end{document}